\documentstyle[12pt]{article}

\begin{document}
\title{Unparticle as a field with continuously distributed mass}
\author{N.V.Krasnikov  
  \\
INR RAN, Moscow 117312}

\maketitle
\begin{abstract}

We  point out  that the notion of an unparticle, 
recently introduced  by Georgi,  can be interpreted as a 
particular case of a field  with continuously distributed mass
considered in ref.\cite{14}.
We also point out that the simplest  renormalizable extension of the 
$SU_c(3)\otimes SU_L(2)\otimes U(1)$
Standard Model is the extension with the replacement of the $U(1)$ 
gauge propagator  $\frac{1}{k^2} \rightarrow \frac{1}{k^2} 
+ \int _0^{\infty}\frac{\rho(t)}{-t+k^2 +i\epsilon}dt$ 
with $\int _0^{\infty}\rho(t)dt < \infty  $.

\end{abstract}

\newpage

Recently, Georgi \cite{1} made an interesting observation that a 
nontrivial scale invariant  sector of scale dimension $d_U$ might 
manifest itself at low energy as a nonintegral number $d_U$ of invisible 
massless particles, dubbed unparticle $U$ with untrivial phenomenological 
implications. \footnote{See also some implications in Refs.  
\cite{2}- \cite{13}.}

In this note     
we  point out  that the notion of an unparticle, 
 introduced by Georgi  can be interpreted as a 
particular case of a field  with continuously distributed mass \cite{14}.
We also point out that the simplest  renormalizable extension of the 
$SU_c(3)\otimes SU_L(2)\otimes U(1)$
Standard Model is the extension with the replacement of the $U(1)$ 
gauge propagator  $\frac{1}{k^2} \rightarrow \frac{1}{k^2} 
+ \int _0^{\infty}\frac{\rho(t)}{-t+k^2 +i\epsilon}dt$ 
with $\int _0^{\infty}\rho(t)dt < \infty$.

Let us start with N  scalar fields \cite{14}
$\phi_k(x)$
with masses $m_k ~(k = 1,2,...N)$.
For the field 
$\Phi_{int}(x, m_k,c_k, N) = \sum_{k = 1}^{N}c_k\phi_k(x)$  free propagator
 has the form 
\begin{equation}
D_{int}(k^2, m_k, c_k,N) = \sum_{k = 1}^{N} \frac{|c_k|^2}
{(k^2 - m^2 +i\epsilon)} = \int _0^{\infty}\frac{\rho(t, c_k, m_k, N)}
{k^2 -t+i\epsilon}dt   \,,
\end{equation}
where the spectral density is $\rho(t, c_k, m_k, N) = 
 \sum_{k =1}^{N} |c_k|^2 \delta(t - m^2_k)$.
In the limit $ k \rightarrow \infty$ 
$\rho(t, c_k, m_k, N) \rightarrow \rho(t)$ and the propagator 
 $ D_{int}(k^2, m_k, c_k,N) \rightarrow D_{int}(k^2) =    \int _0^{\infty}
\frac{\rho(t)}{k^2 -t+i\epsilon}dt$ \cite{14}.
For instance, for $m^2_k = m^2_0 +  \frac{k}{N} \Delta^2  $ and 
$|c_k|^2 = \frac{1}{N}$ we find that the limiting spectral 
density is $\rho(t) = \frac{1}{\Delta^2}
\theta (t - m^2) \theta(m^2 + \Delta^2 -t)$. 
For the limiting spectral density $ \rho(t) \sim t^{\delta-1}$ we find that 
the propagator $D_{int}(k^2) \sim (k^2)^{\delta - 1}$ that corresponds to the 
case of unparticle propagator. In other words, for the limiting 
spectral density  $ \rho(t) \sim t^{{\delta}-1}$ the field 
$\phi_{int}(x, \rho(t)) = lim_{N \rightarrow \infty } 
\Phi_{int}(x, m_k,c_k, N)$
can be interpreted as unparticle.
\footnote{The interpretation of the unparticle as a tower of massive 
particles 
was also proposed in ref.\cite{15}}

 It should be stressed that the 
limiting field $\phi_{int}(x, \rho(t))$ can be 
interpreted as a field describing 
scalar particle with continuously distributed mass \cite{14}. 
Moreover  we believe it is important to 
consider possible experimental consequences fot arbitrary 
spectral density $\rho(t)$ but not only for $\rho(t) \sim t^{-\delta}$
corresponding to unparticle case.
One can introduce 
the selfinteraction Lagrangian of the field  $\phi_{int}(x, \rho(t))$ in 
standard way as 

\begin{equation}
L_{int}(\phi_{int}(x,\rho(t)) = -\lambda (\phi_{int}(x,\rho(t)))^4
\end{equation}

For finite  $\int _0^{\infty}\rho(t)dt $
the asymptotics of  propagator  $D_{int}(k^2) \sim \frac{1}{k^2}$ 
and the model (2)  is renormalizable one. It should 
be noted that for Georgi noninteracting scalar unparticle 
effective Lagrangian is 
\begin{equation}
L_{unp} = \frac{1}{2}\partial_{\mu} \phi (-\partial^{\mu}
\partial_{\mu})^{-\delta}  \partial^{\mu}\phi \,.
\end{equation}

The Lagrangian  $L_{tot} = L_{int} + L_{unp}$  has 
generalized scale invariance \cite{16} and all ultraviolet divergent integrals 
can  be made finite by subtraction of infinities at zero external 
momentum. 

For the Standard Model based on 
$SU_c(3)\otimes SU_L(2)\otimes U(1)$ gauge group there are several ways 
to generalize it by the introduction of the fields with 
continuously distributed mass. Namely, it is possible to introduce new 
scalar field  $\phi_{int}(x, \rho(t))$ with continuously 
distributed mass and introduce the interaction with
standard Higgs doublet field $H(x)$ 
\begin{equation}
 L_{int}(\phi_{int}(x,\rho(t), H(x)) = -\lambda_2 
(\phi_{int}(x,\rho(t)))H^{+}(x)H(x) \,.
\end{equation}
After electroweak symmetry breaking the singlet field  
$\phi_{int}(x, \rho(t))$ will mix with the standard Higgs boson 
that can change drastically \cite{15} Standard Model predictions for 
the Higgs boson search at LHC. 

The generalization to the case of vector fields is the following. Consider 
the Lagrangian
\begin{equation}
L_0 = \sum_{k=1}^{N}[-\frac{1}{4e^2_k}F^{\mu\nu,k}F_{\mu\nu,k} 
+\frac{m^2_k}{2e^2_k}(A_{\mu,k} - \partial_{\nu}\phi_k)^2] \,,
\end{equation}
where $F_{\mu\nu,k} = \partial_{\mu}A_{\nu,k} - \partial_{\nu}A_{\mu,k}$. 
   The Lagrangian (5) is invariant under gauge transformations 
\begin{equation}
A_{\mu,k} \rightarrow A_{\mu,k} +\partial_{\mu}\alpha_k \,,
\end{equation}
\begin{equation}
\phi_k \rightarrow \phi_k + \alpha_k \,.
\end{equation}
For the field $B_{\mu} = \sum_{k=1}^N A_{\mu,k}$ free propagator in 
transverse gauge is 
\begin{equation}
D{\mu\nu}(p) = (g_{\mu\nu} - \frac{p_{\mu}p_{\nu}}{p^2})
(\sum^N_{k=1}( \frac{e^2_k}{p^2-m^2_k})) \,. 
\end{equation}
In the limit $N \rightarrow \infty $
\begin{equation}
D_{\mu\nu}(p)  
\rightarrow (g_{\mu\nu} - \frac{p_{\mu}p_{\nu}}{p^2}) D_{int}(p^2)\,, 
\end{equation}
where 
\begin{equation}
D_{int}(p^2)  = \int _0^{\infty}\frac{\rho(t)}{p^2 - t +i\epsilon}dt
\end{equation}
and $\rho(t) \geq 0$. 
One can introduce the interaction of the field $B_{\mu}$ with  
fermion field $\psi$ in standard way, namely 
\begin{equation}
L_{int} = \bar{\psi}\gamma_{\mu}\psi B^{\mu} \,.
\end{equation}
The simplest generalization of the Standard  Model 
consists in the the replacement of the $U(1)$ gauge 
field propagator 
\begin{equation}
  (g_{\mu\nu} - \frac{p_{\mu}p_{\nu}}{p^2})\frac{1}{p^2} \rightarrow 
(g_{\mu\nu} - \frac{p_{\mu}p_{\nu}}{p^2})D_{int}(p^2) \,.
\end{equation}
This generalization of the 
Standard Model preserves the renormalizability for finite  
$ \int _0^{\infty}\rho(t)dt$ because the ultraviolet asymptotics of 
$ D_{int}(k^2)$ coincides with  free propagator. 
For $\rho(t) \sim t^{\delta -1}$ we reproduce the case of vector unparticle.
For the propagator $D_{int}(p^2) = g^2_1(\frac{1}{{p^2}} + 
\frac{1}{(p^2  - M^2)})$ we obtain generalization of 
the Standard Model with single additional vector field. Current TEVATRON 
experimental bound on $M$ is $M \geq 850~GeV$ \cite{18}. For the 
model with arbitrary $D_{int}(p^2)$ experimental bound will depend 
on the spectral density $\rho(t)$.

It should be stressed that the fields with continuously distributed mass arise 
naturally in n-dimensional field theories \cite{17}.
Consider  five-dimensional scalar field  with the Lagrangian
\begin{equation}
L_5 = \frac{1}{2}(\partial_{\mu} \phi \partial^{\mu}\phi - 
\phi f(-\partial_4^2)\phi) \,,
\end{equation}
where $\mu = 0,1,2,3$. For the Lagrangian $L_5$ free propagator is 
\begin{equation}
D_0 = \frac{1}{k_{\mu}k^{\mu} - f(k^2_4)} \,.
\end{equation}
For the field $\phi(x_{\mu}, x_4 = 0)$ 
propagator is proportional to $\int^{\infty}_{-\infty}\frac{dk_5}{
k_{\mu}k^{\mu} - f(k^2_5)}$  that corresponds to the case of the 
field with continuously distributed mass. 
The fact that unparticle can be interpreted as a result of the 
compactification of 5-dimensional space-time for a model with $AdS_5$ 
metric was mentioned also in ref.\cite{15}. The difference of our model and 
the model proposed in \cite{15} is that we use 5-dimensional Lagrangian  (13) 
which explicitly violates five-dimensional Poincare group. 
The Lagrangian (13) is invariant only under 4-dimensional Poincare group.

This work was supported by the Grant RFBR 07-02-00256, 
I am indebted to a referee for pointing out to me a reference \cite{15} and 
critical remarks.

\end{document}